\documentstyle{article}

\let\a=\alpha\let\d=\delta
\let\e=\epsilon\let\g=\gamma
\let\th=\theta\let\l=\lambda
\let\m=\mu\let\n=\nu
\let\s=\sigma

\let\S=\Sigma

\newcommand{\be}{\begin{equation}}
\newcommand{\ee}{\end{equation}}
\newcommand{\bea}{\begin{eqnarray}}
\newcommand{\eea}{\end{eqnarray}}

\newcommand{\del}{\partial}

\newcommand{\nbox}{{\,\lower0.9pt\vbox{\hrule \hbox{\vrule height 0.2 cm \hskip 0.2 cm \vrule height 0.2 cm}\hrule}\,}}

\begin{document}
\begin{titlepage}
\begin{center}
\vskip .2in \hfill \vbox{
    \halign{#\hfil         \cr
           hep-th/0208085 \cr
            UCI-TR 2002-30\cr
           }  
      }   
\vskip 0.5cm
{\large \bf Comments on D-branes in flux backgrounds}\\
\vskip .2in {\bf Arvind Rajaraman} \footnote{e-mail address:
arajaram@uci.edu}
\\
\vskip .25in {\em Department of Physics,University of California,
Irvine, CA 92697.  \\} \vskip 1cm
\end{center}
\begin{abstract}
We construct supergravity solutions for D-branes in nontrivial
flux backgrounds. We revisit the issue of charge quantization in
this framework, and show that in these backgrounds, charge need
not be quantized. We also show in a particular example that the
semiclassical description of branes produces integral charges.
 \end{abstract}
\end{titlepage}

\section{Introduction}

If monopoles are present in a theory, electric charge is
quantized. This observation of Dirac is fundamental to any
quantum-mechanical theory. Yet there appears to be a violation of
this basic principle in a class of string theories.

The apparent violation occurs for D-branes in the presence of a
background NSNS three-form flux. A variety of arguments show that
in these theories, D-branes have a charge of the form $\sin(\pi
j/k)$, where $j,k$ are integral. In this paper, we shall reexamine
the arguments for this strange behaviour, and show how they are in
fact compatible with Dirac's argument. Other proposals have been
made to resolve this problem: we discuss the connection of our
solution to these other suggestions.

We first describe the boundary state construction of D-branes on
$S^3$ \cite{Cardy:ir}, where this behaviour was first noted. A
semiclassical description of these branes was given in \cite{BDS}
and it was argued that the semiclassical description also produced
nonquantized charges. Taylor \cite{Taylor:2000za} later argued
that in fact, terms not considered by \cite{BDS} made the charge
integral. We review all these arguments in the following sections.

We then construct a supergravity description of such branes, which
provides a complementary approach.  We show that all the features
of the semiclassical description are reproduced by the
supergravity solution. The supergravity approach in general also
allows us to generalize to the case of RR backgrounds, which
cannot be treated in the boundary state approach.

In this paper, we focus on   the case of M-branes in a $S^4$
background, which as we show displays all the features of D-branes
in $S^3$.
 The supergravity description tells us the field
configuration outside the M-branes. The charge of the M-brane is
associated with a massless gauge field $A$ whose long distance
falloff is governed by the harmonic equation
 \bea
 {1\over \sqrt{G}}\del_A(\sqrt{G}G^{AB}\del_B A)=0
 \eea
 Here $G$ is
the asymptotic metric.

The falloff therefore crucially depends on the form of the
asymptotic geometry. It turns out that the falloff is much faster
in the case of asymptotically $AdS$ geometries as compared to
asymptotically flat geometries. Thus there is no long range
interaction between a fundamental charge and a monopole in
asymptotically $AdS$ spaces, and the Dirac argument does not lead
to charge quantization. This is very similar to the proposal of
\cite{BDS} for resolving the charge quantization issue. However,
we also argue that in fact, the semiclassical description produces
quantized charges, in agreement with Taylor \cite{Taylor:2000za}.

We close with a discussion of related issues. We show that
deriving the flat space S-matrix from the AdS/CFT correspondence
must deal with issues raised by the existence of the fractionally
charged states. Also we comment on the issue of charge
conservation in these theories.

\section{The boundary state construction}

D-branes in flat space are described as endpoints of open strings.
In more general theories, it is necessary to use the more abstract
language of boundary states to describe them. In this language,
D-branes are described in terms of their overlap amplitudes with
closed string states. The connection to open strings is made
through Cardy's condition described below.

Boundary states are typically hard to construct, but the special
case of D-branes in $AdS_3 \times S^3$ with background NS-NS flux
can be analyzed exactly, since the $S^3$ factor can be represented
by the exactly solvable $SU(2)$ WZW model. Boundary states in this
theory were constructed in \cite{Cardy:ir}. We review this
construction here.

One first tries to solve the equation \bea
(J^a+\bar{J}^a)|I\rangle=0\label{isheq}\eea The solutions to this
equation are the Ishibashi states $|I_n\rangle$, which are
constructed by taking any primary $\phi_n$ and summing the
normalized states in the Verma module. More precisely, start with
a primary $|\phi^j\rangle$ (by this we mean a state which is
annihilated by all lowering operators $J^a_n$, $n>0$). Define the
Ishibashi state \cite{Ishibashi:1988kg}
 \bea
 |I^j\rangle=M_{IJ}^{-1}J_{-I}\bar{J}_{-J}|\phi^j\rangle
 \label{ish}
 \eea
 Here $I,J$ are ordered strings of indices
 $(n_1,a_1)\dots(n_r,a_r)$
 and $J_I=J_{n_1}^{a_1}\dots J_{n_r}^{a_r}$
 The normalization is defined by
 \bea
 M_{IJ}=\langle\phi^j|J_IJ_{-J}|\phi^j\rangle
 \eea
$M_{IJ}$ is invertible for any module. It is straightforward to
show that the states (\ref{ish}) satisfy the equation
(\ref{isheq}).

There is therefore a 1-1 correspondence between the Ishibashi
states and the primaries. In the $SU(2)$ WZW theory at level $k$,
the primaries are labelled by an index $j$ that runs from 0 to
$k$. Hence there are $k+1$ Ishibashi states.

 One then constructs
Cardy states, which are linear combinations of Ishibashi states
satisfying Cardy's condition. Cardy's condition is that the
modular transform of the overlap of any two boundary states should
be interpretable as an open string partition function. This
ensures that the D-branes can be reinterpreted as endpoints of
open strings.

The Cardy states in $SU(2)$ WZW theory are found in terms of the
S-matrix $S_a^b$ of the theory to be \cite{Cardy:ir}
  \bea
 |B_j\rangle=\sum_n{S_j^n\over \sqrt{S_0^n}}|I^n\rangle
\eea

 The charge of the state $|B_j\rangle$ is given by its overlap with the
lowest primary, and is hence proportional to \bea
 Q_j={S_j^0\over \sqrt{S_0^0}}\propto \sin{\pi(2j+1)\over k+2}
 \eea

The charge of the brane labelled by the integer $j$ is therefore
not an integer multiple of the brane labelled by $j=1$. Hence the
charges are not quantized.

\section{ The semiclassical approach}
\label{semsec}
 We now want to turn to a semiclassical description
of these branes, following \cite{BDS}.

 The boundary states that we have described
satisfy $(J^a+\bar{J}^a)|B\rangle=0$. It was shown in
\cite{Alekseev:1998mc} that the semiclassical description of such
branes is that of D2-branes wrapping conjugacy classes in $S^3$.
This motivates the study of the dynamics of  branes wrapping such
conjugacy classes. As we now see, the phenomenon of fractional
charges can be reproduced by this  calculation, as shown by
Bachas, Douglas, and Schweigert \cite{BDS}.

The background metric of the $S^3$ is taken to be
 \bea
ds^2=k\a'(d\th^2+\sin^2\th d\chi^2+\sin^2\th \sin^2\chi
d\phi^2)
 \eea and the three form flux is
  \bea
 H_{\th\chi\phi}=2k\a'\sin^2\th\sin\chi
 \eea

The conjugacy classes in $S^3$ are 2-spheres. Examples of such
conjugacy classes are the hypersurfaces $\th={\rm constant}$.
 We should therefore  consider the
dynamics of a D2-brane wrapped on such a hypersurface.

 The brane
interacts with the background curvature and 3-form flux through
the
 Born-Infeld
action
 \bea
   S=T\int d^3\s \sqrt{\det(G_{ab}+B_{ab}+F_{ab})}
 \eea

Here $G_{ab}=G_{\m\n}\del_aX^\mu\del_bX^\nu$ and
$B_{ab}=B_{\m\n}\del_aX^\mu\del_bX^\nu$ are respectively the
pullbacks of the spacetime metric and two form field to the
D-brane worldvolume. $F_{ab}$ is the worldvolume field strength.

In general, we can take the D2-brane to have a nontrivial gauge
field on its worldvolume. Flux quantization implies that $\int
d\s^\mu d\s^\nu F_{\mu\nu} =n$ with n integral.

For a brane wrapped on the conjugacy class $\th={\rm constant}$,
we can choose a static gauge of the form \bea \s^0=t\quad
\s^1=\phi\quad \s^2=\chi \eea We then find an energy functional
for the field $\th(\s^0,\s^1,\s^2)$, which is
 \bea
 E_n(\th)=4\pi k\a'T\left(\sin^4\th+(\th-{\sin2\th\over 2}-{\pi n\over k})^2\right)^{1/2}
 \eea

The equation of motion has a static solution for $\th={\pi n\over
k}$. Therefore, there are stable solutions where the 2-brane wraps
a hypersurface of constant $\th={\pi n\over k}$ for any $n$.

It can further be shown that these branes are BPS, and have a mass
and induced D0-brane charge which is proportional to $\sin({2\pi
n\over k})$. We refer to \cite{BDS} for details. Hence we again
find the phenomenon of nonquantized charges.

\subsection{An objection}

It is however rather difficult to conclude on the basis of this
semiclassical computation that the brane charge is fractional.
This was pointed out by Taylor \cite{Taylor:2000za}, who suggested
that bulk interactions could compensate for the nonintegral
charge, by providing another potential source for the brane
charge.

Let us denote the gauge field coupling to the D0-brane charge as
$A_\m$ and  the gauge field coupling to the D2-brane charge as
$C_{\m\n\rho}$. Furthermore, denote the field strength
corresponding to $C_{\m\n\rho}$ by $F_{\m\n\rho\s}$.

There is a term in the type II string action of the form
$F_{\m\n\rho\s}A^\m H^{\n\rho\s}$, where $H^{\n\rho\s}$ is the
NS-NS flux. The equation of motion for $A_\m$ therefore contains
an extra contribution proportional to $F_{\m\n\rho\s}
H^{\n\rho\s}$. Now the presence of the D2-brane implies that
$F_{\m\n\rho\s}$ must be nonzero, and of course we have a
nontrivial NSNS flux in the background. So this contribution is
nonzero, and presumably should be added to the charge obtained by
the semiclassical computation in the previous subsection.

Taylor further argued in \cite{Taylor:2000za} that the  total
charge obtained by this procedure was indeed integral (see also
\cite{Alekseev:2000jx}).

This might suggest that there is no issue here; the charges of all
the branes are in fact integral. However, this cannot be the
complete solution. The boundary state construction is {\it exact}
in string theory; in other words, it sums up all the corrections
including the supergravity interactions mentioned above. Therefore
since the boundary state has a fractional charge, the total charge
including all bulk interactions is fractional, whether or not one
trusts the semiclassical computation of \cite{BDS}. We shall
return to this issue in a later section.

\section{ The supergravity description}

\subsection{Notation}
We now turn to the supergravity description of these branes.

 There
are several reasons why such a description is useful. The
supergravity approach is complementary to the boundary state and
the semiclassical approaches, since it can be applied to cases
where the boundary state cannot at present be constructed, and to
cases where NS5-branes or M-branes are involved. We will also be
able to address the issue of
 charge quantization in this
 approach.

Now, the brane in $AdS_3 \times S^3$ is a special case of a brane
in the background of another  brane or set of branes. The
supergravity solution for this brane should be a special case of
the more general case of supergravity solutions for intersecting
branes.
 The method for constructing  such solutions for
 brane intersections was
 given in \cite{Rajaraman:2000ws,Rajaraman:2000dn}; we will review this
 construction below. We shall then show that we can indeed find
 expanded brane solutions, similar to those  of the previous sections,
  as a special
 case of these solutions.

 For concreteness we will examine a special case where
 an M2-brane is ending on a set of M5-branes.
 We shall orient the M5-branes along the $x_0,x_1,x_2,x_3,x_4,x_5$
directions, and the M2-branes will be  oriented along the
$x_0,x_1,x_6$ directions and end on the M5-brane.

For notational purposes, we shall label the coordinates $x_0,x_1$
by $x_i$. The coordinates $x_2,x_3,x_4,x_5$ will collectively be
labelled $x_a$, and the coordinates $x_7,x_8,x_9,x_{10}$ will
collectively be labelled $x_\a$.

 In the limit when the number $N$ of M5-branes
 is large, the supergravity solution for the M5-branes goes over
to the metric
 \begin{eqnarray}\label{adsmetric}
\lefteqn{ds^2\equiv G_{\m\n}dX^\m dX^\n=}\quad \quad\nonumber
 \\
 & &={r\over N^{1/3}}\left(-dx_0^2+dx_1^2+dx_a^2\right)
 +{N^{2/3}\over r^2}\left(dx_6^2+dx_\a^2\right)\nonumber
\\
& & ={r\over
 N^{1/3}}\left(-dx_0^2+dx_1^2+dx_a^2\right)
 +{N^{2/3}\over r^2}\left(dr^2\right)+N^{2/3}d\Omega_4^2
 \end{eqnarray}
 which is the metric of $AdS_7 \times S^4$. Here $d\Omega_4^2$ is
 the metric of $S^4$.

  The M2-branes (which are oriented along
$x_0,x_1,x_6$) in this new metric are oriented along $x_0,x_1,x_r$
 and occupy a point in  the $S^4$.

 The considerations of \cite{BDS} suggest that the semiclassical description
 of
 the M2-branes should actually be that of M5-branes oriented
 along $x_0,x_1,x_r$ and
 wrapping a $S^3$
 hypersurface
 in $S^4$. To stabilize this extended brane configuration, there
 should be a
nontrivial
 worldvolume flux on the M5-brane. (Just as
 a nontrivial 2-form field strength can stabilize a brane wrapped on
 $S^2$, a nontrivial 3-form field strength can stabilize
 the M5-brane wrapped on $S^3$.) We could therefore
 perform a semiclassical calculation analogous to to verify this.
 Instead, we will explicitly derive this from the
 supergravity description of
 the branes.

We should therefore look at the supergravity solution for
M2-branes ending on M5-branes.
 The general approach to constructing such solutions was given in
 \cite{Rajaraman:2000ws,Rajaraman:2000dn}.
 One is attempting to find solutions to the Killing
 spinor equations, which for
 instance in the case of 11-D supergravity, is
\bea
 \del_\m\e-{1\over 4}\omega_\m^{ab}\g_{ab}\e+{i\over
 288}(\g_\m^{abcd}-8e_\m^a\g^{bcd})G_{abcd}\e=0
 \eea
Here $G_{abcd}$ is the four-form field strength of 11-dimensional
supergravity.

\subsection{Ansatz for the Killing equation}
We will solve the Killing equation by assuming an ansatz
for the Killing spinor.
 We  take the ansatz for the Killing spinor  to be \bea
\e=(g_{00})^{1/4}\e_0\eea where $\e_0$ is a constant spinor.

Furthermore, there are constraints on the constant spinor $\e_0$
since the supersymmetry is partly broken.  Each brane imposes a
constraint on the spinor $\e_0$. The way this happens is that each
brane is associated with a projector $P_i$ satisfying
$P_i*P_i=P_i$. For example the presence of a M5-brane oriented
along $x_0,x_1,x_2,x_3,x_4,x_5$ imposes the equation $P_1\e_0=0$
where $P_1=(1+i\g_{678910})/2$. Similarly the M2-brane oriented
along the $x_0,x_1,x_6$ imposes the equation $P_2\e_0=0$ where
$P_2=(1+i\g_{016})/2$.

For a system containing several branes we should impose all the
separate projection equations on $\e_0$. In the case we are
considering, we therefore take the constraints on the spinor
$\e_0$  to be $P_1\e_0=P_2\e_0=0$.


After imposing all these constraints,  the Killing spinor
equations reduce to a set of algebraic equations, which can be
used to solve for the field strengths and the metric.

\subsection{Solution}
For the case of M2-branes ending on M5-branes, the
solution was found in \cite{Rajaraman:2000ws}. The metric was
found to be of the form
 \bea
  ds^2=\l^{-2/3}H^{-1/3}(-dx_0^2+dx_1^2)+\l^{1/3}H^{-1/3}dx_a^2
   ~~~~~~~~~~~~~\nonumber
  \\
  +\l^{-2/3}H^{2/3}(dx_6+\phi_a dx^a)^2+\l^{1/3}H^{2/3}dx_\a^2
  \eea

where we must impose the constraint \bea
\del_6(H\phi_a)=\del_aH\label{taueq1}\eea

 The field strengths are given by (all
the indices are  world indices) \bea G_{0167}=\del_7({1\over
\l})~~~~ \qquad \qquad \qquad \qquad \qquad \qquad \qquad
G_{0147}=\del_7({\phi_4\over \l})
\\
G_{0164}=\del_4({1\over \l})-\del_6({\phi_4\over
\l})~~~~~~~~~~~~~~~~~~
 \nonumber
\\
G_{2345}=H^{-1}\del_6\l+\phi_a\del_6\phi_a \qquad\qquad
\qquad\qquad \qquad G_{2357}=-\del_7\phi_4
\\
G_{2356}=-\del_6\phi_4~~~~~~~~~~~~~~~~~~~~~~~~~~~~~~
 \nonumber
 \\
G_{78910}=\del_6(H\l)~~~~~~~ \qquad \qquad \qquad
 \qquad \qquad \qquad G_{68910}=-\del_7H
\\
G_{48910}=-\del_7(H\phi_4)~~~~~~~~~~~~~~~~~~~~~~~~
 \nonumber
 \eea

Hence there are two independent functions $\l,H$ describing the
solution. This is as it should be since we have two types of
branes.

\section{Interpretation of the solution}
\subsection{Source terms}
What is the physical meaning of this solution? To answer this
question, we must look at the various sources, which can be found
by looking at the Bianchi identities and the equations of motion.

We first define the quantities $P, Q$ through \bea\label{taueq2}
P=-\del_a\phi_a+H^{-1}\del_6\l+\phi_a\del_6\phi_a\eea
and\bea\label{taueq3} \del_6
Q=\del_6^2(H\l)+\del_\a^2H~~~~~~~~~~\eea

$Q$ also satisfies\bea
\del_aQ=\del_6\del_a(H\l)+\del_\a^2(H\phi_a) \eea

The nontrivial Bianchi identities are then found to be
 \bea
\del_6G_{2345}+{\rm cyclic}=\del_6P  \label{eq:B62345}
\\
\del_7G_{2345}+{\rm cyclic}=\del_7 P \label{eq:B72345}
\\
\del_4G_{78910}+{\rm cyclic}=\del_4 Q \label{eq:B478910}
\\
\del_6G_{78910}+{\rm cyclic}=\del_6 Q  \label{eq:B678910}\eea

while the nontrivial equations of motion are \bea
\del_A(\sqrt{g}G^{A235})+{1\over
2.(24)^2}\e^{235abcdefgh}G_{abcd}G_{efgh}= {1\over
\l}(\del_4 Q-\phi_4\del_6Q)~~~~~~~~~~\label{eq:E235}\\
\del_A(\sqrt{g}G^{A8910})+{1\over
2.(24)^2}\e^{8910abcdefgh}G_{abcd}G_{efgh}={1\over \l}\del_7
P~~~~~~~~~~~~~~~~~~~~~~~~\label{eq:E8910}
\\
\del_A(\sqrt{g}G^{A017})+{1\over
2.(24)^2}\e^{017abcdefgh}G_{abcd}G_{efgh}=-H\del_7P~~~~~~~~~~~~~~~~~~~~~~\label{eq:E017}
\\
\del_A(\sqrt{g}G^{A014})+{1\over
2.(24)^2}\e^{014abcdefgh}G_{abcd}G_{efgh}=\phi_4\del_6
Q-\del_4Q~~~~~~~~~~~~~~~\label{eq:E014}
\\
\del_A(\sqrt{g}G^{A016})+{1\over
2.(24)^2}\e^{016abcdefgh}G_{abcd}G_{efgh}=
P\del_6(H\l)-~~~~~~~~~~~~~~~~~\nonumber
\\
\phi_a\del_a
Q+\int(dx_6)(\del_a^2Q-\del_\a^2(HP))\label{eq:E016} \eea

The RHS of the Bianchi identities and the equations of motion
should be interpreted as sources like M5-branes and M2-branes.

For instance if there are no M2-branes, and $N$ M5-branes oriented
along $x_0,x_1,x_2,x_3,x_4,x_5$, then only the RHS of
(\ref{eq:B678910}) should be nonzero. In this case we have
 \bea
\del_6Q=N\d(x_6)\d(x_\a) \qquad P=\phi_4=0\qquad \l=1\eea

\subsection{Interpretation of sources}
We now present the  interpretation of these sources in the more
general case when we have both M5-branes and M2-branes. In this
more general case, both $P$ and $Q$ are nonzero, and this produces
 source terms for all the Bianchi identities and the equations of
motion. To simplify the analysis, we will focus on the limit where
the number $n_2$ of M2-branes is much smaller than the number $N$
of M5-branes, so that the picture of branes in the metric
(\ref{adsmetric}) is still valid.

Some of the sources are easy to interpret. As we have already
seen, the RHS of (\ref{eq:B678910}) is the density of M5-branes
oriented along $x_0,x_1,x_2,x_3,x_4,x_5$. Similarly the RHS of
(\ref{eq:E016}) is the density of M2-branes oriented along
$x_0,x_1,x_6$.

However, it would seem strange to interpret the source term on the
RHS of (\ref{eq:E014}) as the density of M2-branes oriented along
$x_0,x_1,x_4$. Luckily, there is a better interpretation. There is
a 2-form gauge field on the M5-brane worldvolume, which couples to
the spacetime gauge fields through the term $\int d^6\s
A^{(3)}h^{(3)}$. A nonzero flux $h_{014}$ thus provides a source
for the field $A_{014}$ i.e. we can set the right hand of
(\ref{eq:E014}) to be \bea h^{014}=\phi_4\del_6 Q-\del_4Q \eea
Similarly the right hand side of (\ref{eq:E235}) can be identified
with $h^{235}$, which is equal to $h_{014}$ since the 3-form is
self dual.

The source terms on the RHS of (\ref{eq:B62345},\ref{eq:B72345})
are very interesting. The source in (\ref{eq:B62345}) has a
natural interpretation as M5-branes oriented along
$x_0,x_1,x_7,x_8,x_9,x_{10}$. Similarly, the source in
(\ref{eq:B72345}) has a natural interpretation as M5-branes
oriented along $x_0,x_1,x_6,x_8,x_9,x_{10}$.

Such sources  represent a M5-brane occupying a hypersurface in the
\newline
$x_0,x_1,x_6,x_7,x_8,x_9,x_{10}$ directions. If we go to the
coordinates in the metric (\ref{adsmetric}), we see that the brane
occupies the $x_0,x_1$ directions and traces a hypersurface in the
$x_r\times S^4$ directions. If we wish to ensure that the brane
occupies a hypersurface in $S^4$ alone, we must take \bea
r\del_rP=x_6\del_6+x_\a\del_\a P=0 \eea  The brane now occupies
the $x_0,x_1,x_r$ directions and a $\theta=\theta_0$ hypersurface.
This provides a connection between our solution and the
semiclassical description of section (\ref{semsec}).

Remarkably, the correspondence can be made even more explicit.
 A flux $h_{8910}$ on the worldvolume of this brane
produces the source in equation (\ref{eq:E8910}), and its
self-dual partner sources equation (\ref{eq:E017}). In the metric
(\ref{adsmetric}), we find that this is exactly a
 constant worldvolume field
 strength  on the M5-brane. This makes the correspondence with
 section (\ref{semsec}) precise.

The remaining source terms can be understood as deformations of
the M5-branes.

We thus see that the branes discussed in \cite{BDS} represent a
special subclass of the general supergravity solution given above.

\subsection{Long distance behaviour} Now we look at the long
distance behaviour of the field configuration. This requires the
usual assumption that the sources are localized in a finite
region. Hence in the long distance region, we can set $P=Q=0$.

We will expand around the background geometry of the M5-brane
 \bea
H=H_0\qquad \l=1\qquad \phi_a=0
 \eea
To leading order the  perturbations
 \bea
  \d H=H-H_0\qquad \d\l=\l-1\qquad \d\phi_a=\phi_a
  \eea
satisfy the system of equations
(\ref{taueq1},\ref{taueq2},\ref{taueq3}) with $P=Q=0$. The
solution to this system of equations is
 \bea
 \d H=\del_6^2\tau \qquad H_0\phi_a=\del_6\del_a \tau\qquad \d
 \l=\del_a^2\tau
 \eea
where $\tau$ satisfies
 \bea
 (\del_6^2+\del_\a^2+H_0\del_a^2)\tau=0\label{taueq}
 \eea
 This  is more recognizable as the harmonic equation
 \bea
  \label{harm}
 {1\over \sqrt{G}}\del_A(\sqrt{G}G^{AB}\del_B \tau)=0
 \eea
 where $G$ is the unperturbed metric for the M5-branes defined in (\ref{adsmetric}).

 Hence $\tau$ satisfies the equation for a massless field in
 $AdS_7\times S^4$. (Note that this is different from \cite{BDS}, where it
 was suggested that $\tau$ was massive.)

 This is the solution far away from the branes. More generally,
 the sources $P,Q$ will appear on the RHS of equation
 (\ref{taueq}). The general solution for
 $\tau$ is then
 \bea
 \tau=\int dx' H(x,x')\rho(x')
 \eea
 where $H(x,x')$ is the Green's function satisfying
\bea
 {1\over \sqrt{G}}\del_A(\sqrt{G}G^{AB}\del_BH(x,x')
 )=\d(x-x')
 \eea
and $\rho(x')$ is a source density.

\section{Discussion}

\subsection{Charge quantization}

 We now
examine the behaviour of $\tau$. On shorter distance scales, much
less than the radius of the $AdS_7$, the space is locally flat (if
$N>>1$) and so $\tau$ has the usual flat space behaviour.

 For
distances larger than the radius of the $AdS_7$, the curvature
becomes important. The long-distance behaviour depends on the form
of $H_0$. If $H_0\rightarrow 1$ at long distances, then
asymptotically $\tau$ again has the flat space behaviour.

But we can also consider the near-horizon limit of the M5-brane,
where $H_0= {r_0^3\over r^3}$. Then the curvature does not vanish
asymptotically, and the equation for $\tau$ does not go over to
the flat space equation.

We would like to understand the long distance behaviour of $\tau$
in this case. Unfortunately, in this case, we cannot solve for the
Green's function explicitly. We instead note that the equation
(\ref{harm}) is now the harmonic equation for a massless field in
$AdS_7\times S^4$. We can therefore expand the fields  in
harmonics on the $S^4$.

The higher harmonics are massive and hence fall off exponentially
in $AdS_7$. The long distance behaviour is therefore dominated by
the lowest harmonic,  and hence we  only need to consider the
lowest harmonic on the $S^4$. Note that this is similar to the
discussion in section 2, where the charge is defined as the
overlap of the boundary state with the lowest harmonic.

This harmonic (which we denote $\tau_0$) satisfies the equation
 \bea
 (y^2+2y)\del_y^2 \tau_0+ (7y+7)\del_y \tau_0=0
 \eea
where we have defined
\bea
 y={\sqrt{rr_0}\over 2}
 \left(
   {(\sqrt{r}-\sqrt{r_0})^2\over rr_0} +
 {(x^a-x^a_0)^2\over 4N}\right)
 \eea
Here $x^a_0,r_0$ are constants labelling the position of the
branes.

We see that for large $y$ the solution has the behaviour
$\tau_0\sim y^{-6}$, while for small $y$, the solution has the
behaviour $\tau_0\sim y^{-5/2}$. Clearly the falloff is much
steeper for large $y$.

Note that for  small $y$, the falloff is appropriate for a
massless field in 6+1 dimensions. (This is consistent with the
idea that for short distances $AdS_7$ resembles $R^{6,1}$). For
large $y$ the falloff is faster, so $\tau$ behaves as if it has an
effective mass.

The long distance behaviour of the gauge fields is controlled by
the behaviour of $\tau$ through the supergravity solution. Since
$\tau$ has a steep falloff at long distances, we correspondingly
find a steep falloff of the gauge field. In effect, the massless
gauge field appears to get an effective mass due to the background
curvature. Due to this behaviour, we find that there is no
requirement of charge quantization for asymptotically $AdS$
spaces. The fields fall off too quickly to have a long distance
interaction. This explanation is similar to that of \cite{BDS}.

To summarize, in the case of asymptotically flat geometries, we
have the usual quantization condition. In the case of
asymptotically $AdS$ spaces, there is no quantization condition,
since the fields fall off too fast.

\subsection{Integral charges}
There are other puzzles raised by this situation, though. We are
claiming that for very large but finite $N$, the semiclassical
picture produces quantized charge. Yet, in the limit $N=\infty$,
the semiclassical solution becomes nonquantized. A local observer
in $AdS_7\times S^4$ would then be able to determine the
asymptotic geometry from local measurements. This is therefore a
gross violation of locality.

The natural objection to this is that in this limit, the
nonquantized spectrum, which is of the form $Q_n=\sin\left
(n\pi\over N\right)$ goes over to $Q_n=\left (n\pi\over N\right)$,
a quantized spectrum. But clearly, if we scale $n$ with $N$, this
is not the case. Therefore the problem remains.

The solution to this puzzle is that demanding locality actually
requires quantized charges. This is easily seen from the
supergravity solution. The field strength on the brane is a
constant. It is obtained from the RHS of equation
(\ref{eq:E8910}). The shape of the brane in $S^4$, on the other
hand, is given by (\ref{eq:B72345}). Clearly to linear order in
perturbations, they are proportional with a fixed constant of
proportionality (in fact, they are exactly proportional even at
the nonlinear level, once the effects of the curved geometry are
taken into account.)

Hence the field strength is constant on the brane,
 and the
constant does not depend on the size of the brane. Since the total
flux is quantized, this implies that the volume of the brane is
quantized. This further implies that the mass and charge of the
brane are quantized.

Note that this argument does not depend on the asymptotic
geometry. It only requires that the local observer, sees branes
that are interpretable as flat space branes in the large $N$
limit. Such an observer must necessarily see quantized charges.
This is furthermore in agreement with Taylor's argument in
\cite{Taylor:2000za}, where he argued that the semiclassical
picture a la \cite{BDS} should always result in quantized charges.

There is no inconsistency with the boundary state analysis. We
have already shown that charges need not be quantized in
asymptotically curved spaces. The only issue is whether there is a
semiclassical description of the boundary states. Our analysis
indicates that there is not such a description. More precisely, it
would appear that the boundary states constructed in section 2
correspond to states in the semiclassical description  where the
flux on the brane is not quantized.

\subsection{On the flat space S-matrix}

 The AdS/CFT correspondence provides a nonperturbative definition
 of
 string theory on $AdS$ spaces. It has been proposed
 \cite{Polchinski:1999ry,Susskind:1998vk} that this
 indirectly also
 provides a nonperturbative definition
 of
 string theory on flat space by considering a limit where the
 interaction region is small compared to
 the scale of the $AdS$ space. Intuitively, the curvature
 of the space should be invisible in this limit, and we should
 reproduce scattering in flat space.

 However, if the BPS spectrum in the $AdS$ space differs from the
 flat space BPS spectrum, then the interactions will not be
 reproduced. The existence
 of the fractionally charged objects might then lead to a
 breakdown of the map from $AdS$ space to flat space.

This is most easily exhibited in the case of $AdS_3\times
\S^3\times K3$, when the K3 is taken to have a size $l_s$ (the
string length), while $AdS_3\times \S^3$ has a large radius $kl_s$
with $k$ large. If we consider interactions occurring in a small,
string-scale sized local region of $AdS_3\times \S^3$ , then we
might expect to reproduce scattering in $R^{5,1}\times K3$.

Now in $R^{5,1}\times K3$ there are no bound states of two
D0-branes. In $AdS_3\times \S^3\times K3$, on the other hand there
does exist such a state. This has a charge ${k\over
\pi}\sin{2\pi\over k}\sim 2$ times the D0-brane charge. The
semiclassical construction of \cite{BDS} provides an explicit
description of another  state, which has charge exactly $2$ times
the D0-brane charge. We know the size of this second state; it is
of order $l_s$.

 Now if either of
these bound states appears in the interaction then it will produce
a pole (or cut) in the $AdS_3\times \S^3\times K3$ correlations
which would not appear in the flat space S-matrix. This would then
indicate that the flat space S-matrix would not be obtained as a
limit of the $AdS$ scattering. Since the second state above  has a
small size, of order $l_s$, there is no obvious reason it should
not appear in the interactions. This therefore seems to lead to a
breakdown of the derivation of the flat space S-matrix (at least
in this particular case, but similar issues exist in the cases
with greater supersymmetry).

It would be very interesting to see if there is a way that this
issue is resolved, or whether it is really the case that the flat
space S-matrix cannot be obtained from the $AdS$ theory.

\subsection{Charge quantization}

 Finally, we comment on the issue of charge quantization. Since we
 now
 have branes of charge
 ${k\over \pi}\sin{2\pi\over k}< 2$, one can ask if two branes of
charge 1 can decay to a bound state with the charge given above.

We wish to argue that this in fact cannot occur. The charge is
still coupled to a massless gauge field. The consistency of the
gauge transformation implies charge conservation. In other words,
the lack of charge conservation leads to a gauge anomaly. This
would appear to be fraught with difficulties.

It was argued in \cite{Fredenhagen:2000ei} that in fact one could
find a transition between a set of separated D0-branes and a bound
state which is of the form described in section (\ref{semsec}). We
have argued that in fact this new state is also integrally
charged, and hence charge conservation is maintained. (A similar
argument was given in \cite{Alekseev:2000jx}). Presumably there is
no transition between separated D0-branes and a fractionally
charged state.

\section{Summary and Conclusions}

We have presented a supergravity solution for branes in
$AdS_7\times S^4$ which displays the fact that branes which are
naively pointlike in the $S^4$ are in fact extended objects on the
$S^4$. The existence of a constant flux on the worldvolume was
found as a direct consequence of the supergravity solution. The
construction works similarly for branes in other $AdS$ geometries.

Using this construction, we were able to resolve several confusing
issues regarding these branes. We showed that the charges of these
branes were not quantized despite the fact that they were coupled
to a massless gauge field. This was because the curvature of the
$AdS$ space led to an effective mass for the gauge field at long
distances.

Furthermore, we were able to show that the semiclassical
description \cite{BDS} of this brane always produced integral
charges. This is most directly seen in the construction we have
given. Other calculations require a careful accounting of bulk
charges and brane charges, which is automatically taken care of
here.

Finally, we have commented on several other issues, including the
problems of reproducing the flat space S-matrix, and charge
conservation in these theories.



\end{document}